\documentclass[12pt, preprint]{aastex}
\usepackage{amsmath}
\usepackage{graphicx}
\usepackage{natbib}

\usepackage[colorlinks=true,citecolor=blue,breaklinks=true,linktocpage=true]{hyperref}
\bibpunct{(}{)}{;}{a}{}{,} 
\usepackage[switch,pagewise]{lineno}
\usepackage{xspace}
\usepackage{hyperref}

\newcommand{\alfven}{Alfv\'en\xspace}
\newcommand{\unit}[1]{\ensuremath{\,\mathrm {#1}}}  
\renewcommand{\d}{\ensuremath{\text{d}}}  

\newcommand{\figref}[1]{Figure~\ref{#1}}
\newcommand{\secref}[1]{Section~\ref{#1}}
\renewcommand{\eqref}[1]{Equation~\ref{#1}}

\usepackage{color}

\shorttitle{Fast waves in structured plasmas}
\shortauthors{Yuan et al.}

\begin{document}

\title{Evolution of fast magnetoacoustic pulses in randomly structured coronal plasmas}
\author{D. Yuan\altaffilmark{1,2,3}}
\email{Ding.Yuan@wis.kuleuven.be}
\author{D.~J. Pascoe\altaffilmark{4}}
\author{V.~M. Nakariakov\altaffilmark{4,5,6}}
\author{B. Li\altaffilmark{1}}
\email{bbl@sdu.edu.cn}
\and
\author{R. Keppens\altaffilmark{2}}

\altaffiltext{1}{Shandong Provincial Key Laboratory of Optical Astronomy and Solar-Terrestrial Environment, Institute of Space Sciences, Shandong University,Weihai, 264209, China}
\altaffiltext{2}{Centre for mathematical Plasma Astrophysics, Department of Mathematics, KU Leuven, Celestijnenlaan 200B bus 2400, B-3001 Leuven, Belgium}
\altaffiltext{3}{Key Laboratory of Solar Activity, National Astronomical Observatories, Chinese Academy of Sciences, Beijing, 100012}
\altaffiltext{4}{Centre for Fusion, Space and Astrophysics, Department of Physics, University of Warwick, Coventry CV4 7AL, UK}
\altaffiltext{5}{School of Space Research, Kyung Hee University, Yongin, 446-701, Gyeonggi, Korea}
\altaffiltext{6}{Central Astronomical Observatory of the Russian Academy of Sciences at Pulkovo, 196140 St Petersburg, Russia}

\begin{abstract}
Magnetohydrodynamic waves interact with structured plasmas and reveal the internal magnetic and thermal structures therein, thereby having seismological applications in the solar atmosphere. We investigate the evolution of fast magnetoacoustic pulses in randomly structured plasmas, in the context of large-scale propagating waves in the solar atmosphere. \textbf{We perform one dimensional numerical simulations of fast wave pulses propagating perpendicular to a constant magnetic field in a low-$\beta$ plasma with a random density profile across the field.} Both linear and nonlinear regimes are considered. We study how the evolution of the pulse amplitude and width depends on their initial values and the parameters of the random structuring. A randomly structured plasma acts as a dispersive medium for a fast magnetoacoustic pulse, causing amplitude attenuation and broadening of the pulse width. After the passage of the main pulse, secondary propagating and standing fast waves appear in the plasma. Width evolution of both linear and nonlinear pulses can be well approximated by linear functions; however, narrow pulses may have zero or negative broadening. This arises because a narrow pulse is prone to splitting, while a broad pulse usually deviates less from their initial Gaussian shape and form ripple structures on top of the main pulse. A linear pulse decays at almost a constant rate, while a nonlinear pulse decays exponentially. A pulse interacts most efficiently with a random medium which has a correlation length of about half of its initial pulse width. The development of a detailed model of a fast MHD pulse propagating in highly structured medium substantiates the interpretation of EIT waves as fast magnetoacoustic waves. Evolution of a fast pulse provides us with a novel method to diagnose the sub-resolution filamentation of the solar atmosphere.
\end{abstract}

\keywords{Sun: atmosphere --- Sun: corona --- Sun: oscillations --- magnetohydrodynamics (MHD) --- methods: numerical --- waves}

\section{Introduction}
\label{sec:intro}

Magnetohydrodynamic (MHD) waves play an important role in the formation and dynamics of the solar atmosphere.
They may contribute significantly to coronal heating \citep[see reviews by][]{klimchuk2006,parnell2012}, 
  and solar wind acceleration and heating \citep[see e.g.,][]{suzuki2006,li2007,li2008,ofman2010}. 
\textbf{In a uniform magnetised plasma there are four MHD eigenmodes, disregarding the directions of propagation:  \alfven, fast and slow magnetoacoustic modes, and entropy mode, see \citet{goedbloed2004}.} Only the fast magnetoacoustic mode can propagate in a direction perpendicular to the magnetic field, and therefore can experience a wealth of plasma structuring. Plasma structuring modifies MHD wave propagation significantly, causing the effects of wave-guiding, dispersion, linear coupling, resonant absorption and phase-mixing \citep[see, e.g.,][]{edwin1983,vandoorsselaere2008,sakurai1991,heyvaerts1983, goedbloed2004,goedbloed2010}. 
Various MHD eigenmodes of \textit{regularly} structured plasma non-uniformities are well understood theoretically, and are confidently detected at different heights of the solar atmosphere \citep[see reviews by][]{nakariakov2005rw,ofman2005,demoortel2012,pascoe2014b}. 
MHD wave theory in regular plasma non-uniformities created a solid basis for the rapid development of wave-based
   plasma diagnostics technique -- MHD coronal seismology \citep[see][for a recent review]{stepanov2012}.  
MHD seismology has been successfully applied to measuring the magnetic field \citep{nakariakov2001}, 
   or quantifying transverse loop structuring \citep{aschwanden2003}, \alfven transit times \citep{arregui2007}, polytropic index \citep{vandoorsselaere2011}, thermal conduction coefficient \citep{ofman2002c}, the magnetic topology of sunspots \citep{yuan2014cf,yuan2014lb}, and the magnetic structure of large-scale streamers \citep{chen2010,chen2011}.

On the other hand, MHD waves in a \textit{randomly} structured plasma are less commonly studied. 
Before the commissioning of high-resolution EUV observations of the solar corona, the main interest in the effect of 
   random structuring on MHD waves was connected with interstellar medium and sunspot applications. 
In particular, \citet{valley1971,valley1974} demonstrated analytically that linear MHD waves experience attenuation in
   a medium with a stochastic magnetic field. \citet{li1987} considered analytically the propagation of an \alfven 
   wave package in a medium with time-dependent short-wavelength random density irregularities and found that the \alfven waves transferred power into compressive modes which dissipate and heat the plasma. 
\citet{lazzaro2000} further developed the analytical theory of linear MHD waves in randomly structured media, 
   and applied it to a specific case of a \lq\lq two-phase plasma\rq\rq\ that consisted of randomly spaced 
   regions with two alternative \alfven speeds. 
Scattering of acoustic waves that interact with a bundle of field-aligned fibrils was comprehensively studied
   by \cite{bogdan1985, bogdan1987}. \citet{keppens1994} studied multiple scattering of acoustic waves in
   fibril spaghetti sunspots and showed that impinging $p$-modes suffer enhanced damping due to resonant
   absorption in successive scatters~\citep[also see][]{labonte1993, ryutova1993b,ryutova1993a}.  Fast magnetohydrodynamic modes of multi-fibril bundles, representing coronal loops, were considered by \cite{diaz2005}. 
\citet{gruszecki2006} modelled the effect of random structuring on the period, decay time and 
    amplitude of standing kink modes of a plasma cylinder. 
Sausage oscillations of a randomly multi-shelled coronal loop were numerically modelled by \cite{pascoe2007} 
    who found that the period of oscillation was insensitive to very fine structuring. 
Random structuring was considered in the context of the damping rate \citep{pascoe2011} and 
   effect of line-of-sight integration \citep{demoortel2012b} for propagating kink waves undergoing mode coupling \citep{pascoe2010}.

An analytical model based on the eikonal formalism was developed for describing linear shear \alfven waves 
   in a medium with a randomly tangled magnetic field \citep{similon1989}. 
It was found that random structuring of the field affects the wave dissipation significantly by phase mixing. 
Nonlinear self-interaction of broadband shear \alfven waves, which effectively constitute a randomly structured
   medium for the propagation of individual spectral components, was modelled in terms of reduced MHD \citep{kleva1992}. 
The efficient dissipation-less damping of the waves was established. 
A similar conclusion was reached by \cite{malara2003}
   who analytically described the propagation of short-wavelength small-amplitude \alfven waves in a randomly structured
   cold compressive plasma in terms of the WKB approximation. 
Enhanced damping of \alfven waves in a plasma with a balanced MHD turbulence was also evident in the incompressible 
   ideal MHD modelling performed by \cite{beresnyak2008}. 
Recently, \citet{tsiklauri2014} modelled numerically the dynamics of linearly polarised \alfven waves in
   a randomly tangled magnetic field, and found that the wave evolution is strongly affected by the random structuring.

The discovery of global coronal waves, also known as EIT, EUV or coronal Moreton waves, 
   or coronal bright fronts \citep{thompson1999}, stimulated further interests in understanding the interaction of
   MHD waves with a randomly structured coronal plasma. 
These waves are usually triggered by flares or coronal mass ejections and propagate for
   several hundred megameters as EUV, soft X-ray and radio emission intensity disturbances
   at speeds of several hundred kilometers per second \citep[see][for recent reviews]{warmuth2011,gallapher2011,patsourakos2012}.
EIT waves are usually interpreted as fast magnetoacoustic waves \citep[see][for a discussion]{wang2000,wu2001,wu2005,ofman2002}, 
   while alternative interpretations have not been ruled out \citep[see e.g.][]{delannee2000,chen2005,patsourakos2012}. 
However, the deceleration, profile broadening, amplitude attenuation \citep{warmuth2004a,warmuth2004b,long2011} and
   nonlinear steepening \citep{veronig2010, shen2012} of EIT waves appear to be consistent with a fast wave nature. 
Therefore, one could model the interaction of EIT waves with bundles of solar coronal loops as fast magnetoacoustic
   wave propagation through a randomly structured medium. Wave modification by perpendicular structuring of coronal
   plasma is demonstrated by the propagation of kink waves \citep{vandoosselaere2008b}. 
In the context of EIT waves, 
   it is important to consider wave pulses rather than harmonic waves. 
The nonlinear effects connected with the finite amplitude of the waves should also be taken into account. 
\citet{nakariakov2005} studied formation of a dispersive oscillatory wake in a fast magnetoacoustic pulse
   propagating through a randomly structured medium. \citet{murawski2001} investigated possible decrease in the phase speed of EIT wave in terms of fast MHD wave propagation in a cold plasma with random structuring.

In this study, we use a one-dimensional ideal MHD model to simulate the propagation of fast magnetoacoustic pulses perpendicular
   to the magnetic field in a randomly structured plasma. 
The structuring is applied across the field direction, and the parametric values are set in accordance with coronal conditions. 
The purpose of our study is to understand amplitude attenuation and broadening of the pulses caused by the random structuring. 
Our model differs from the study by \citet{murawski2001} in a number of ways: 
   a) we use a random structuring of a uniform spectrum
      in contrast to a Gaussian spectrum; 
   b) \citet{murawski2001} constructed a random profile by evaluating the \alfven speed directly. 
   This means extra mass is loaded to the system by the structuring in comparison to a uniform plasma. 
   This step naturally leads to the retardation of the main pulse; 
   c) \citet{murawski2001} consider only the linear wave regime in an incompressible (zero-$\beta$) plasma, 
   while we study both linear and nonlinear regimes in typical compressible coronal plasmas. 
This paper is organised as follows; the numerical model is described in \secref{sec:model}. 
The simulation results are presented \secref{sec:result}, followed by our conclusions (\secref{sec:disc}).

\section{Numerical model}
\label{sec:model}

We use the \textsc{MPI-AMRVAC} code \citep{keppens2012,porth2014} to solve a set of normalised ideal MHD equations
   in conservative form using a finite volume method,

\begin{equation}
\frac{\partial \mathbf{U}}{\partial t}+\nabla\cdot\mathbf{F(\mathbf{U})}=0 , \label{eq:cons}
\end{equation}
where
\begin{eqnarray*}
\mathbf{U} =\begin{bmatrix}
                        \rho \\
                      \rho\mathbf{v}\\
                        e \\
                    \mathbf{B}
                   \end{bmatrix},
\mbox{ and }
\mathbf{F} =\begin{bmatrix}
                        \rho \mathbf{v}\\
                     \mathbf{v}\rho\mathbf{v}+\mathbf{I}(p+\mathbf{B}^2/2)-\mathbf{B}\mathbf{B}\\
                      \mathbf{v}(e+p+\mathbf{B}^2/2)-\mathbf{v}\cdot\mathbf{B}\mathbf{B}\\
                       \mathbf{v}\mathbf{B}-\mathbf{B}\mathbf{v}
                           \end{bmatrix}
\end{eqnarray*}
   are the conservative variables and flux functions, respectively.
Furthermore, $\rho$ is the mass density, $\mathbf{v}$ is the plasma velocity, 
   $\mathbf{B}$ is the magnetic field, $e=p/(\gamma-1)+\rho \mathbf{v}^2/2+\mathbf{B}^2/2$ is total energy density, 
   $p$ is the gas pressure, $\gamma$ is the adiabatic index, and $\mathbf{I}$ is a $3\times3$ unit matrix. 
\textbf{The ideal MHD equations ensure adiabatic processes, within which the specific entropy is conserved in the Lagrangian frame, i.e., $\d s/\d t=\d(p\rho^{-\gamma})/\d t=0$}.

The MHD equations (\eqref{eq:cons}) are solved in Cartesian coordinates $(x,y,z)$ by assuming that the physical quantities
   do not vary along the $x$- and $z$-directions ($\partial/\partial x =\partial/\partial z=0$). 
Therefore the \textsc{MPI-AMRVAC} code is configured to a 1.5D MHD setup. 
We use an idealised 1D plasma model with a constant magnetic field and a random density profile perpendicular
   to the magnetic vector. 
This kind of approximation was successfully applied using the \textsc{Lare2D} code \citep{arber2001} to
   modelling 1D MHD processes in the solar atmosphere \citep[e.g.,][]{tsiklauri2004,botha2011}. 
In the following text, we consistently adopt dimensionless quantities.
However, one could restore the dimensionalities by choosing any set of three independent
   variables, e.g., $B_0=10\unit{G}$, $L_0=1000\unit{km}$, $\rho_0=7.978\times10^{-13}\unit{kg\, m^{-3}}$, 
   which are typical values for coronal plasmas. 
The \alfven speed  $V_A=B_0/\sqrt{\rho_0 \mu_0}=1000\unit{km\,s^{-1}}$ (where $\mu_0=4\pi\times10^{-7} \unit{N\, A^{-2}}$ is the magnetic permeability of free space)
   and \alfven transit time $\tau_A=L_0/V_A = 1\unit{s}$ serve as the units for the velocity and time, respectively.

\begin{figure*}[ht]
\centering
\includegraphics[width=0.8\textwidth]{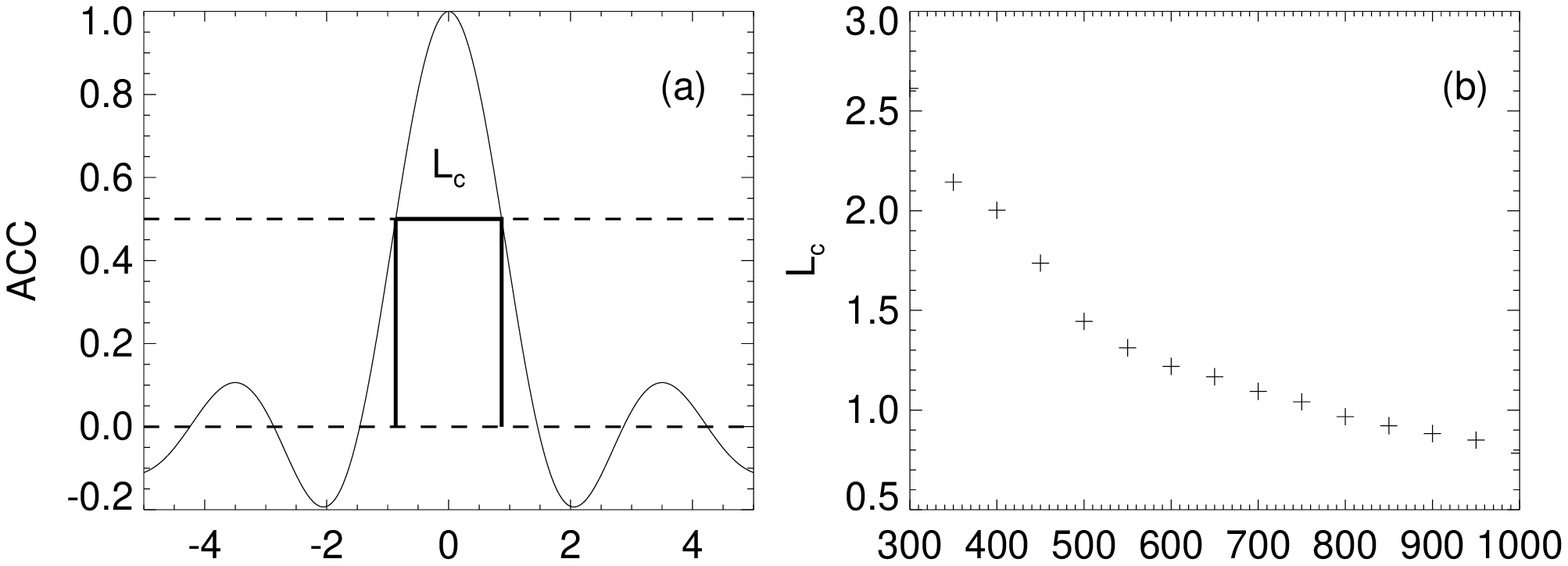}
\caption{
(a) The correlation length $L_c$ of a randomly structured plasma
        defined as the FWHM of the autocorrelation coefficient (ACC) of the profile. 
    The dashed lines present the zero and half-maximum levels of ACC. 
(b) Dependence of the correlation length $L_c$ on the number of harmonics $N$. 
    The correlation length $L_c$ decreases as higher harmonics are added to the density profile.
    }
\label{fig:corrdisp}
\end{figure*}

\begin{figure*}[ht]
\centering
\includegraphics[width=0.8\textwidth]{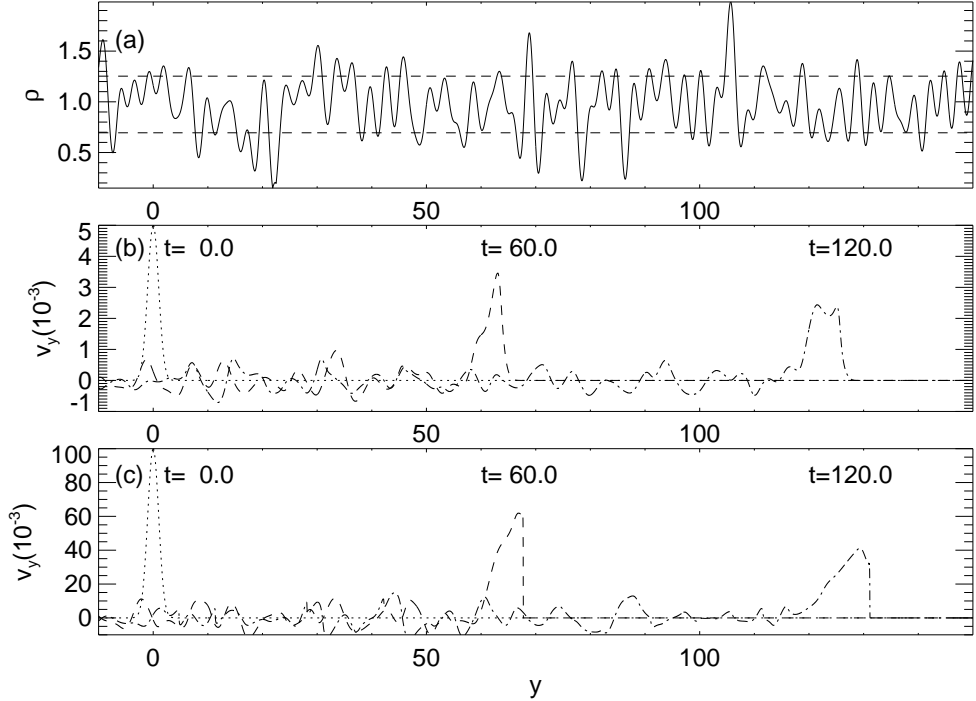}
\caption{
(a): Density profile for a randomly structured plasma with a density contrast $\delta_\rho=0.28$ 
         and a correlation length $L_c=1.46$. 
     The dashed lines mark the density levels at $1\pm\delta_\rho$. 
Small- ((b), $A_0=0.005$) and large-amplitude ((c), $A_0=0.1$) fast magnetoacoustic pulses of width $w_0=2.35$ 
     were launched and demonstrate linear and nonlinear wave effects. 
The snapshots of the pulse evolution are overplotted at $t=0$ (dotted lines), $t=60$ (dashed lines) and $t=120$ (dot-dashed lines).}
\label{fig:wavedisp}
\end{figure*}

The magnetic field was chosen to be constant and in the $x$-direction $\mathbf{B}=\left(1,0\right)$. 
The density profile consists of a set of sinusoidal harmonic perturbations in the $y$-direction added to a constant background:
\begin{equation}
\label{eq:dens}
\rho\left( y \right) = 1+\frac{\Delta}{N}\sum\limits_{i=1}^{N} R_i\sin \left( \alpha k_i y +\phi_i \right),
\end{equation}
where $\Delta$ is a scaling parameter to control the 
   density contrast $\delta_\rho=\left[\int_{0}^{L_y}\left(\rho\left(y\right)-\bar{\rho}\right)^2\d y/L_y\right]^{0.5}$,
   in which $\bar{\rho}=\int_{0}^{L_y}\rho(y)\d y/L_y\simeq1$ is the mean density, and $L_y$ is the size of the numerical domain in the $y$-direction. 
The parameter $k_i=i\pi/L_y$ is the wavenumber of the $i$-th harmonic component. 
The values $R_i$ and $\phi_i$ are the random amplitude and phase of the $i$-th harmonic component, 
   given by a uniform pseudo-random number generator within the ranges $[0,1]$ and [$-\pi$,$\pi$], respectively. 
The number of harmonics $N$ determines the correlation length $L_c$ of the density profile. 
The factor $\alpha$ controls how fast $L_c$ drops off with increasing $N$ and we take $\alpha=1/4$ in our study. 
\figref{fig:corrdisp} shows the definition of the correlation length $L_c$ of the random structuring as the
   full-width-at-half-maximum (FWHM) of the autocorrelation function of the density profile. 
This value can be monotonically controlled by adding more harmonics (i.e., increasing $N$)
   to the density profile (\eqref{eq:dens}). 
\figref{fig:wavedisp}{a} illustrates a typical density profile with $\delta_\rho=0.28$ and $L_c=1.26$ ($N=600$), 
   which implies that about 67\% of the density humps is distributed within 28\% of the mean value in the density profile
   and that they are separated by a distance of $L_c$ on average. \textbf{This scenario could also be viewed as bundles of coronal loops with an average width of $L_c$ and filled with plasmas of random density that fluctuates around the mean value $\bar{\rho}$ with a standard deviation of $\delta_\rho$.}

The computational domain size $L_y$ is set to be $160$. 
We apply open boundary conditions at both sides of the $y$-dimension, so that energy is free to propagate
   out of the simulation domain. 
In practice, this is implemented by zeroth order extrapolation of the conservative variables in the ghost
   cells in combination with the exploited approximate Riemann solver; this also ensures no artificial
   wave reflection at the boundaries. 
In the initial equilibrium the plasma $\beta=0.01$ is enforced over the simulation domain. 
In the initial state total pressure balance $P_\text{total}(y)=\mathbf{B}^2/2+\rho(y)T(y)=(1+\beta)\mathbf{B}^2/2$ is
   ensured by adjusting the temperature profile $T(y)$ accordingly,
   meaning that the temperature profile is also randomly structured.

\textbf{A fast magnetoacoustic pulse is launched by augmenting the equilibrium values with a perturbation in the initial conditions as follows,
\begin{align}
 v'_y (y)  &  = A_0\exp\left[-4\ln{2}\frac{(y-y_0)^2}{w_0^2}\right], \label{eq:wave_v} \\
 B'_x (y) &  =  A_0\exp\left[-4\ln{2}\frac{(y-y_0)^2}{w_0^2}\right], \label{eq:wave_b}  \\
\rho' (y) &  = A_0\exp\left[-4\ln{2}\frac{(y-y_0)^2}{w_0^2}\right] \label{eq:wave_r},
\end{align}
where $v'_y$, $B'_x$, and $\rho' $ are the perturbations of the $y$-component of the velocity, the $x$-component of $\mathbf{B}$ field and plasma density, respectively; $A_0$ and $w_0$ are the initial amplitude and width (FWHM) of the pulse, respectively; and $y_0=0$ is the starting position of the pulse. This specific perturbation ensures that the fast pulse will propagate in one direction.}

\eqref{eq:cons} is solved with the HLL approximate Riemann solver~\citep{harten1983} and
    a three-step Runge-Kutta method in time discretisation~\citep[for details, see][and references therein]{keppens2014, porth2014}. 
HLL is an upwind Godunov-type scheme and uses a constructed flux function with two fastest wave speeds and switch
    the flux accordingly to the orientation of the Riemann fan in the $y$-$t$ space. 
We performed a feasibility study to see how the use of different limiter exploited in the cell
    center-to-edge variable reconstruction behaves in the wave problem by launching a linear
    wave pulse with $A_0=0.005$ and $w_0=0.94$ in a uniform magnetized plasma ($\Delta=0$ in \eqref{eq:dens}). 
We switched between possible limiter options in \textsc{MPI-AMRVAC} and assessed their quality
    by examining how much the wave amplitude is numerically damped by the combined effects of the approximate Riemann solver, 
    time advancing and reconstruction limiter. 
These simulations were run with 9600 fixed grid cells ($\Delta y=0.0167$). 
After travelling a distance of $150$, the fast wave pulse was damped
    by 25.0\% (MINMOD), 9.6\% (WOODWARD), 12.0\% (ALBADA), 10.8\% (MCBETA), 6.1\% (KOREN) 5.8\% (PPM)
    and -5.1\% (cada3\footnote{Note: a negative damping means that CADA3 is too strong a limiter and results in amplification of the pulse.}) \citep[see details in, e.g.,][and references therein]{toth1996,kuzmin2006,keppens2014}. 
    \textbf{We exclude the limiters that cause significant numerical damping, i.e., MINMOD, WOODWARD, ALBADA, MCBETA, and those limiters that have numerical artefacts, i.e. CADA3. Since PPM requires four ghost cells rather than two and much more computation time,} we opt for the KOREN flux limiter \citep{koren1993} and use 12000 fixed grid cells. 
The wave is then damped by about 5\% for $w_0=0.94$, about 1\% for $w_0=1.18-1.41$, and
    less than  0.1\% for $w_0>1.88$ for pulses propagating in a uniform medium. 
Note that these numerical errors are overestimated, if we must take into account that the
    initial pulse setup is not an exact eigenmode solution of the system and that we quantified errors by assuming unchanged pulse shape.
In the uniform medium the pulse does not experience  broadening or develop the oscillatory wake.

\section{Results}
\label{sec:result}

\figref{fig:wavedisp} presents snapshots of the evolution of both a linear and a nonlinear
   fast magnetoacoustic pulse of the same initial width. 
Consistent with previous findings
   discussed in \secref{sec:intro}, the random structuring causes both attenuation and dispersion. 
Nonlinear pulses experience steepening, but the efficiency is reduced by effective attenuation and
   dispersion caused by the random structuring.

\begin{figure}[ht]
\centering
\includegraphics[width=0.5\textwidth]{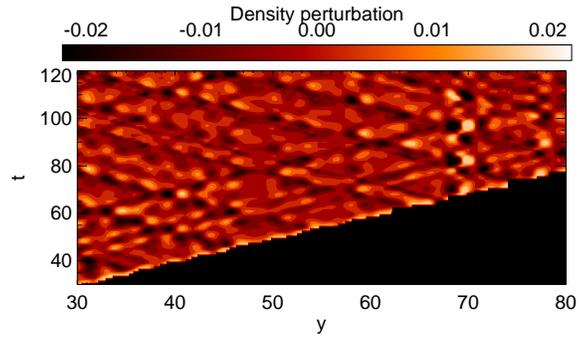}
\caption{
Baseline-difference time-distance plot for the density perturbations $\rho(y,t)-\rho(y,0)$ 
    showing the secondary pulses trapped by a randomly structured plasma after the passage
    of a small-amplitude fast magnetoacoustic pulse. 
This is the same run as in \figref{fig:wavedisp}{b}. 
The blank bottom-right triangular region represents the passage of the main pulse and the associated density perturbation is masked out.
}
\label{fig:rhodisp}
\end{figure}

\figref{fig:rhodisp} shows a typical baseline-difference time-distance plot~\citep[see][for technical details]{yuan2012sm}
   for a fast pulse propagating through a randomly structured medium. 
The main pulse was covered by a dynamic mask (black region), and hence small amplitude density perturbations (coloured region)
   were revealed to illustrate the plasma evolution in the aftermath of the main pulse. 
\figref{fig:rhodisp} reveals two kinds of perturbations. 
The diagonal ridges represent secondary pulses propagating approximately at the average \alfven speed in both directions. 
The average \alfven speed also corresponds to the slope of the line indicating the main pulse, which travels at a speed of
    about unity in our simulations. 
These pulses appear due to reflections from the non-uniformities. Standing features (see, e.g. $y\simeq 69$) that oscillate in time correspond to the eigenmodes of the non-uniformities excited by the main and secondary pulses. The energy that excites secondary propagating and standing features was extracted from the main pulse causing its gradual decay.

\begin{figure*}[ht]
\centering
\includegraphics[width=0.8\textwidth]{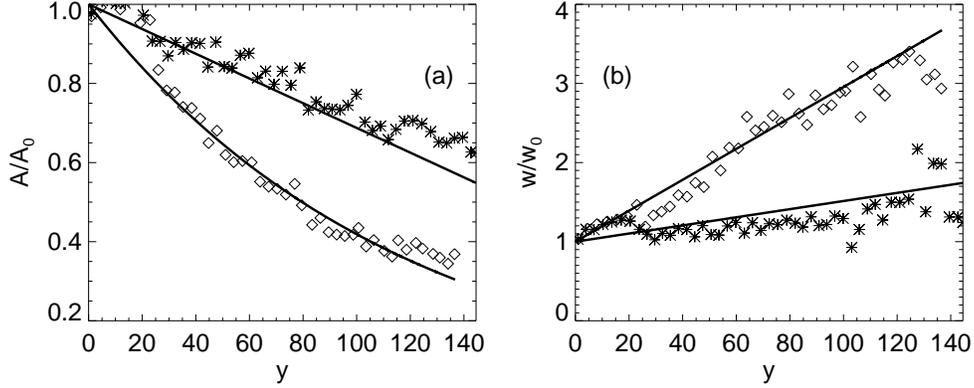}
\caption{
Linear ($A_0=0.005$, asterisk) and nonlinear ($A_0=0.1$, diamond) fast magnetoacoustic pulses
    launched with initial width of $w_0=1.9$ in a randomly structured plasma with a density contrast $\delta_\rho=0.24$
    and a correlation length $L_c=1.3$. Panels (a) and (b) plot the normalised amplitude and width evolution (symbols), 
    respectively. 
An exponential fit is applied to the amplitude evolution of the nonlinear pulse, and other profiles were fitted with a linear function (solid lines).}
\label{fig:decaydisp}
\end{figure*}

To quantify the effect of the structuring we measured the amplitude $A(y)$ and width $w(y)$ of the peak in the velocity perturbation profile for each snapshot.
\figref{fig:decaydisp} shows the evolution of pulse amplitude and width for both a linear (asterisk) and a nonlinear (diamond) wave. 
The width of both linear and nonlinear pulses increase quasi-linearly with propagation distance, so we obtain a characteristic
    width amplification coefficient $K_s$ by fitting the data
    with $w/w_0=1+K_s y$\footnote{We used MPFIT \citep{markwardt2009} package to perform all the fittings, 
    the source code is available at \url{http://purl.com/net/mpfit}.}. 
The amplitude of a linear pulse is damped linearly by the random structuring, while the nonlinear pulse
    decays exponentially, so we fit them with $A/A_0=1-K_a y$ and $A/A_0=\exp(-K^*_a y)$, respectively. 
Note that $\exp(-K^*_a y)\simeq1-K^*_a y$ for $K^*_a\ll1$. 
In the following text and figures, $K_a$ denotes both linear and exponential damping coefficients, where we drop the superscript for the latter.
In the case of \figref{fig:decaydisp}, we obtain $K_a=0.0031$ and $0.0087$ for the linear and nonlinear amplitude evolutions, respectively; 
    while for the width profiles, $K_s=0.0051$ and $0.0196$ were found. 
The coefficients $K_s$ and $K_a$ depend on the initial parameters of the pulse (width $w_0$), 
    and on the parameters of random structuring (density contrast $\delta_\rho$ and correlation length $L_c$). 
To investigate this behaviour we performed a set of parametric studies.

\subsection{Effect of initial pulse width}
\label{sec:width}

\begin{figure*}[ht]
\centering
\includegraphics[width=0.45\textwidth]{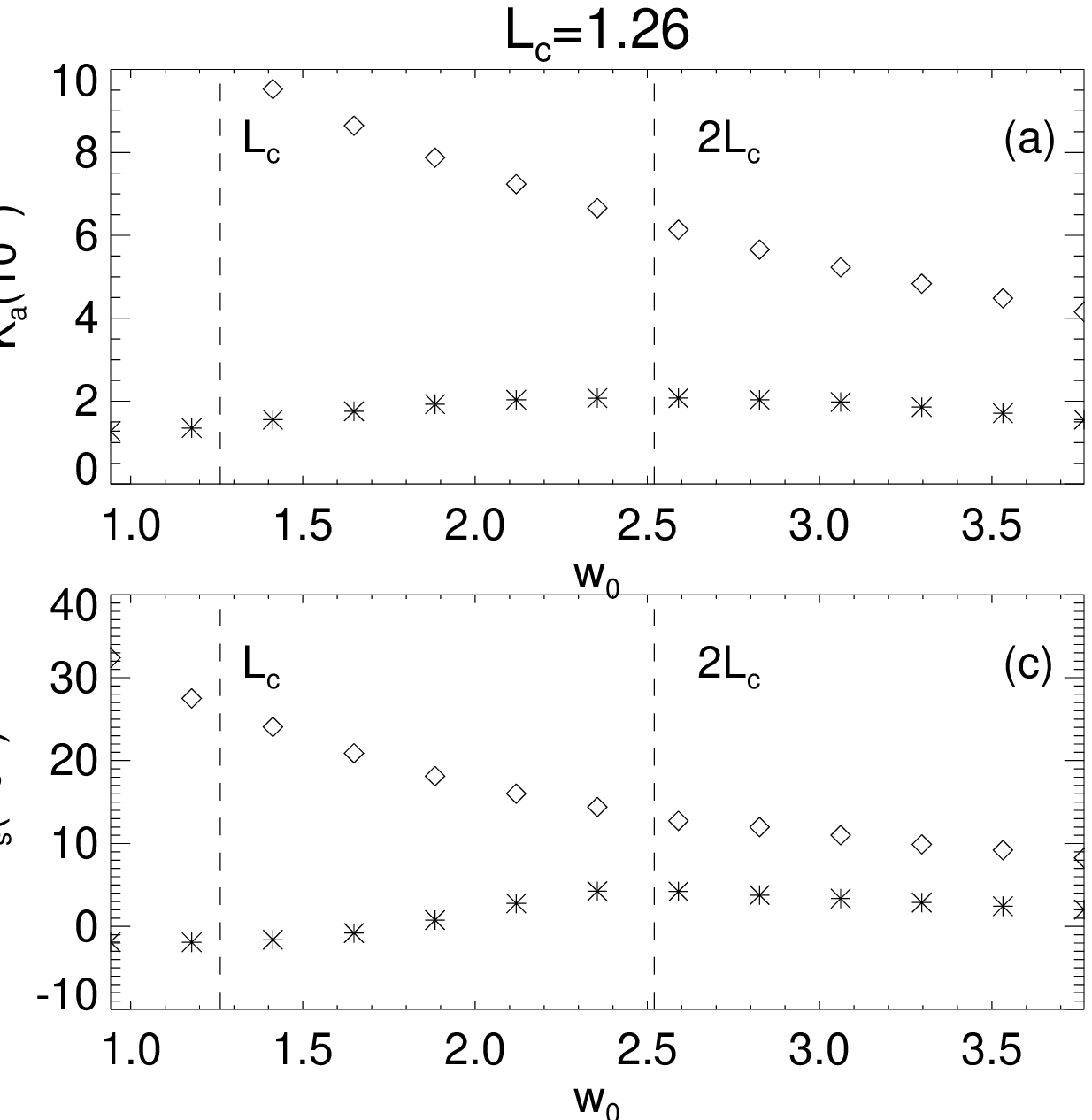}
\includegraphics[width=0.45\textwidth]{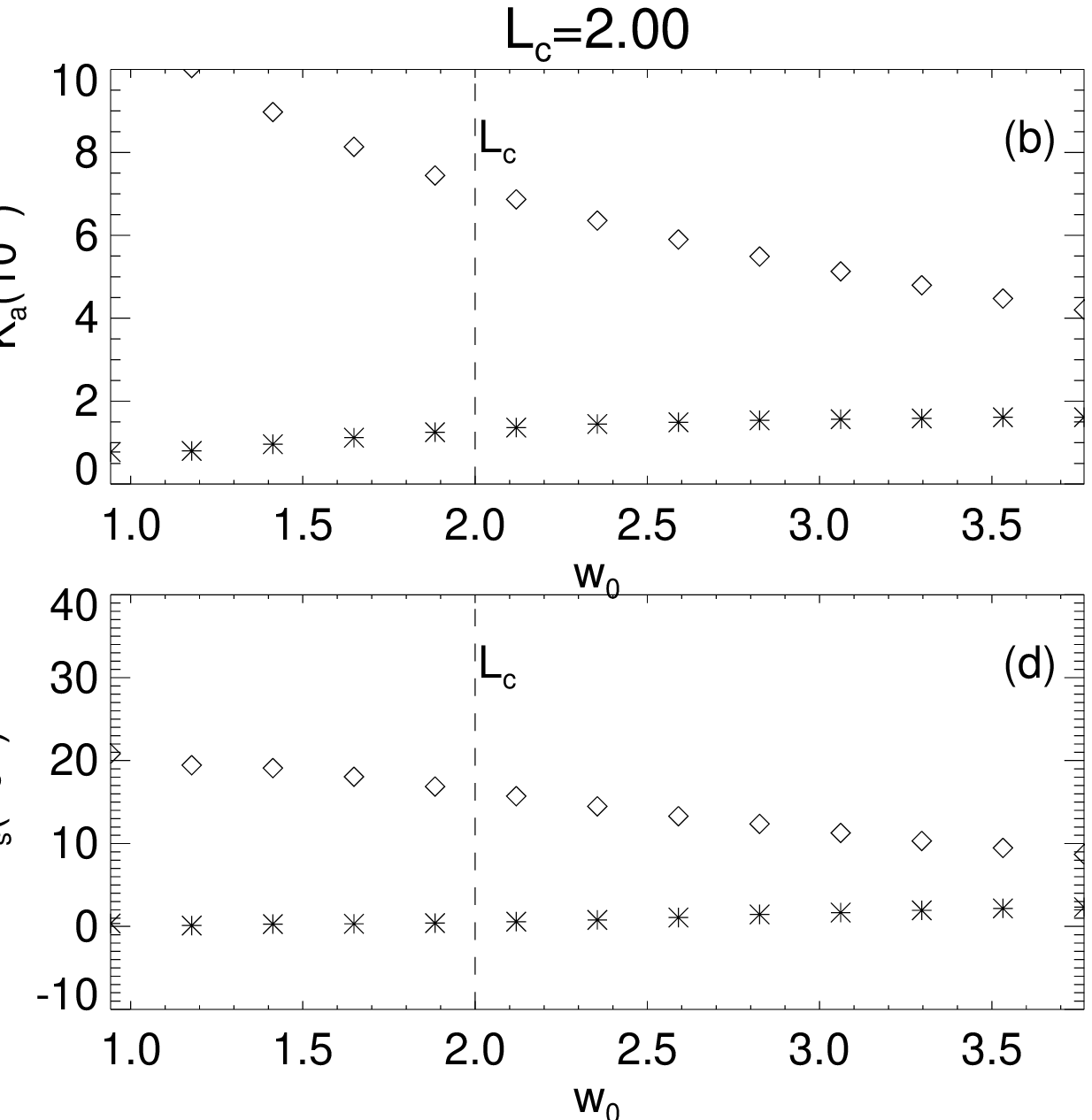}
\caption{
$K_a$ and $K_s$ as a function of the initial pulse width $w_0$ for the linear (asterisks)
   and nonlinear (diamonds) waves in a randomly structured plasma with density contrast $\delta_\rho=0.18$ 
   and correlation length $L_c=1.26$ (a,c) or $L_c=2.0$ (b,d).}
\label{fig:width}
\end{figure*}

\begin{figure*}[ht]
\centering
\includegraphics[width=0.8\textwidth]{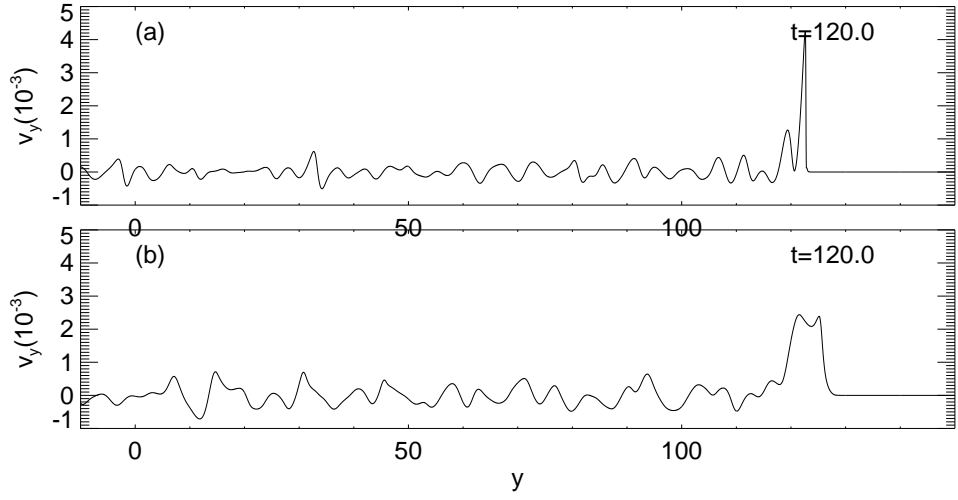}
\caption{
(a) A narrow pulse ($w_0=1.17$) splits into a main pulse and a co-evolving sub pulse. 
(b) A broad pulse ($w_0=2.35$) exhibits sub structuring atop its main pulse.}
\label{fig:finestruct}
\end{figure*}

We first consider the effect of the initial pulse width $w_0$. 
\figref{fig:width} shows that nonlinear pulses ($A_0=0.1$) are attenuated stronger than linear ones ($A_0=0.005$) 
    and their widths also expand faster, i.e., the $K_a$ and $K_s$ values are larger. 
For a linear pulse in a random plasma with $L_c=1.26$ the maximum attenuation and broadening occur for an initial pulse
    width $w_0\simeq2.5$, which is approximately a value of $2L_c$ (\figref{fig:width}{a} and \ref{fig:width}{c}). 
To further verify this point, we launched a set of pulses in a random plasma with $L_c=2.0$ and expect maximum damping
    at $2L_c=4.0$ which is out of the selected range of initial pulse width. 
Indeed, \figref{fig:width}{b} and \ref{fig:width}{d} exhibit no resonant attenuation. 
The turnover is only observed for linear pulses, suggesting that nonlinear steepening and broadening are stronger than
    the resonant attenuation effect. Narrower pulses are typically attenuated more than broader ones. 
We also find that the width broadening coefficient $K_s$ is negative in the vicinity of $w_0\simeq L_c$ 
    for a linear pulse, meaning that the pulse width decreases as it propagates through the random plasma. 
This is due to the fact that a narrow pulse in this case tends to split into a main pulse and a co-evolving pulse, 
    as shown in  \figref{fig:finestruct}{a} which shows two velocity profiles at $t=120$. 
This phenomenon may be considered as the formation of an oscillatory wake behind the main pulse. 
This effect is also seen in the case of a broader pulse, while the main body of the pulse does not
    experience major decrease in that case. 
A sufficiently broad pulse (\figref{fig:finestruct}{b}) retains the structure of a single main pulse
    with multiple ripples on top of the main pulse. \textbf{We did a convergence test and switched to PPM flux limter. In both cases, the spitting of narrow pulses and multiple-ripple structure of wider pulses are well resolved. So we could confidently conclude that these are physical effects caused by the randomly structured plasma alone. }

\subsection{Effect of random structuring density contrast}

\begin{figure*}[ht]
\centering
\includegraphics[width=0.48\textwidth]{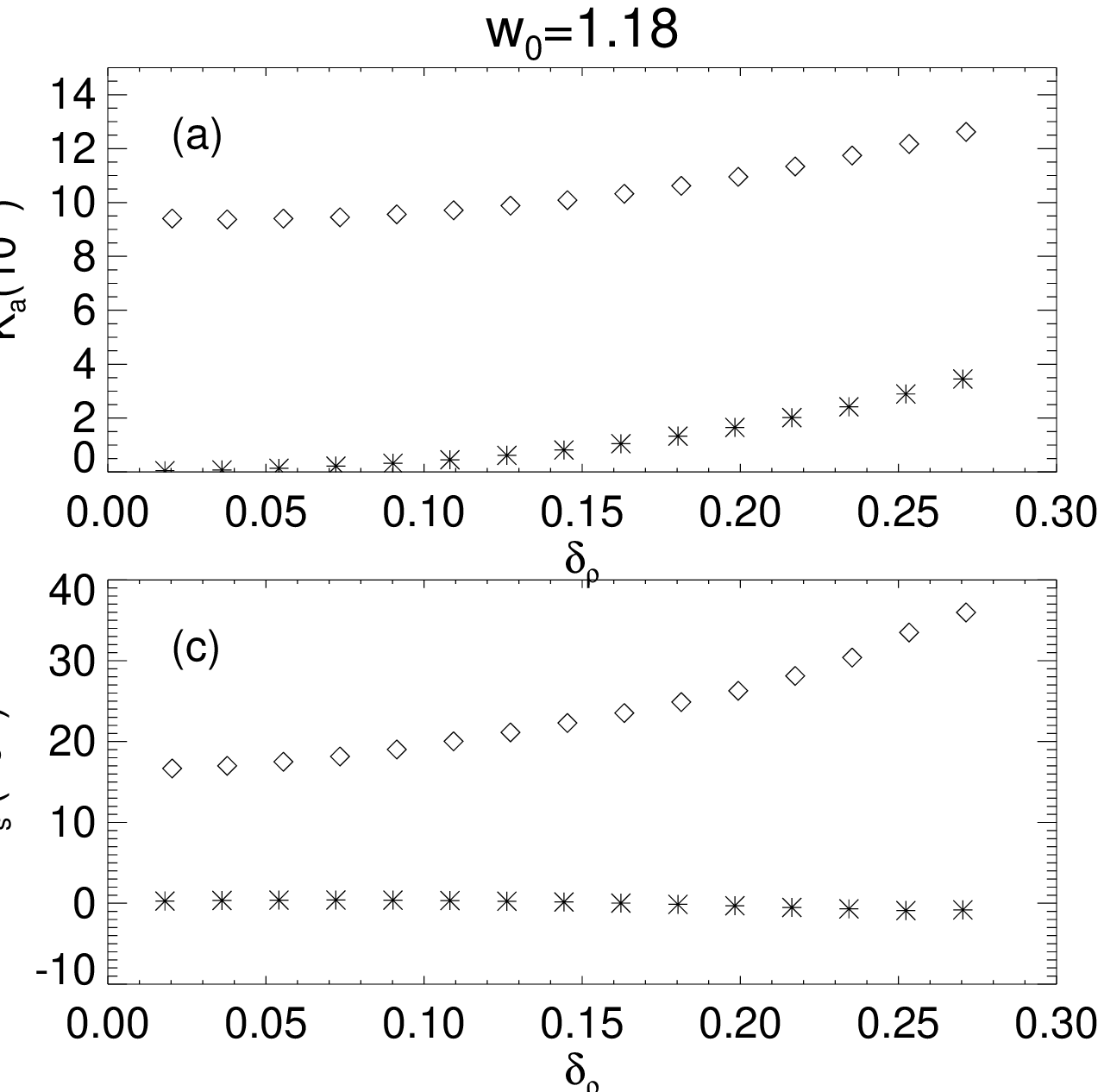}
\includegraphics[width=0.48\textwidth]{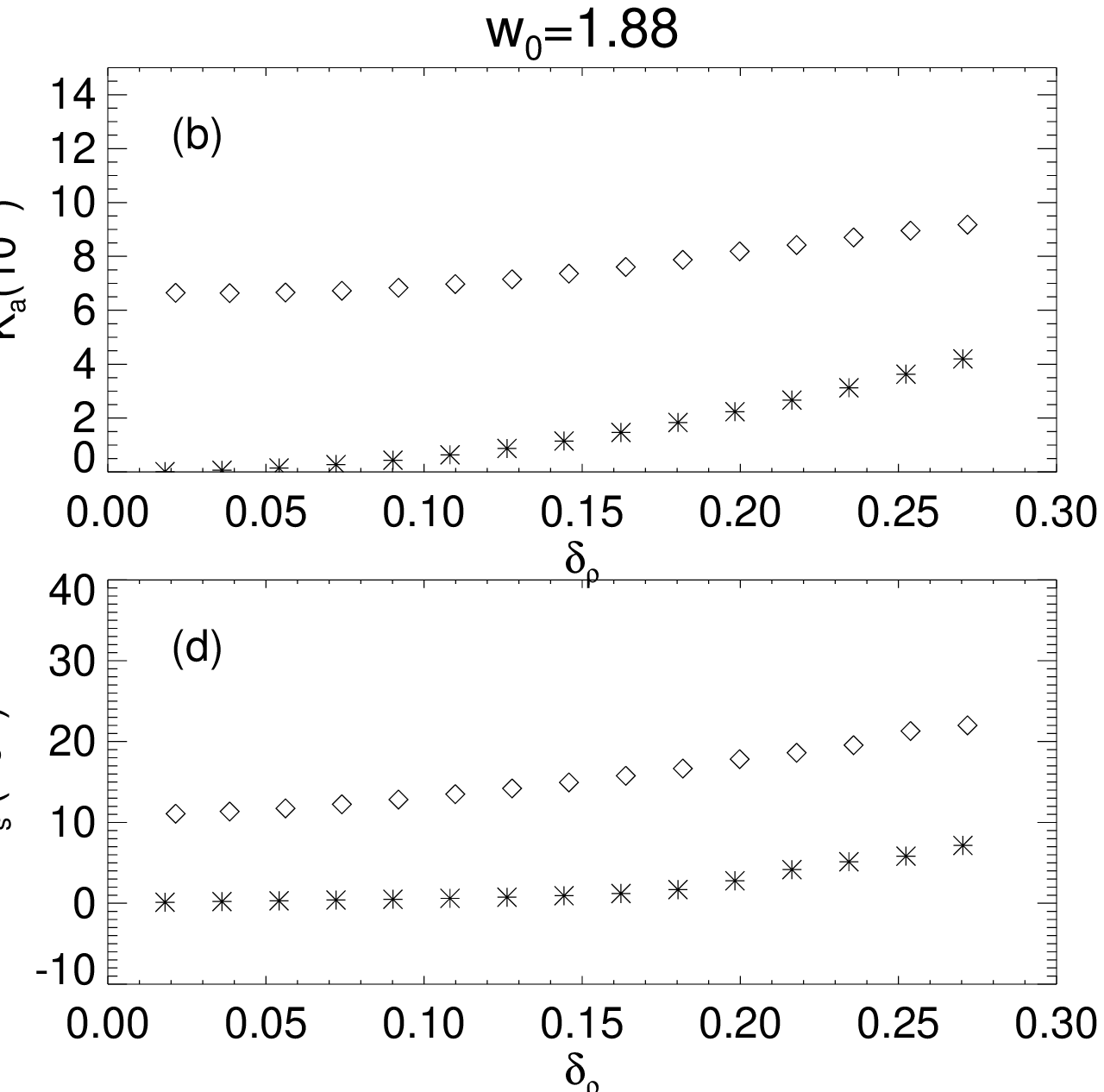}
\caption{
$K_a$ and $K_s$ as a function of the density contrast $\delta_\rho$ of the random structuring
    with $L_c=1.26$ for the linear (asterisks) and nonlinear (diamonds) waves. 
Two sets of pulses were launched with $w_0=1.18$ (a,c) or $w_0=1.88$ (b,d).}
\label{fig:rho}
\end{figure*}

Next we consider the dependence of $K_a$ and $K_s$ on the density contrast $\delta_\rho$. 
\figref{fig:rho}{b} and \ref{fig:rho}(d) show that a larger density contrast typically causes stronger attenuation
    and dispersion for pulses with $w_0$ greater than $L_c$. 
However, in case of narrow pulses (\figref{fig:rho}{a} and \ref{fig:rho}{c}), a higher density contrast encourages
    the splitting of the pulse (see \figref{fig:finestruct}) and so gives zero or negative broadening. 
Therefore, a non-monotonic dependence occurs for $K_s (\delta_\rho)$ in the linear pulse region; 
    small density contrasts not strong enough to cause pulse splitting result in positive broadening, 
    and $K_s$ increases with $\delta_\rho$. 
If $\delta_\rho$ is large enough to cause pulse splitting, $K_s$ decreases with increasing $\delta_\rho$, 
    even to zero or negative values (see \figref{fig:rho}{c}). 
For the pulse amplitude, a higher $\delta_\rho$ usually results in a larger attenuation coefficient $K_a$. 
We note that $K_a$ and $K_s$ exhibit a steeper gradient in the linear wave regime, suggesting that nonlinear waves
    are less sensitive to the random structuring.

\subsection{Effect of random structuring correlation length}

\begin{figure*}[ht]
\centering
\includegraphics[width=0.48\textwidth]{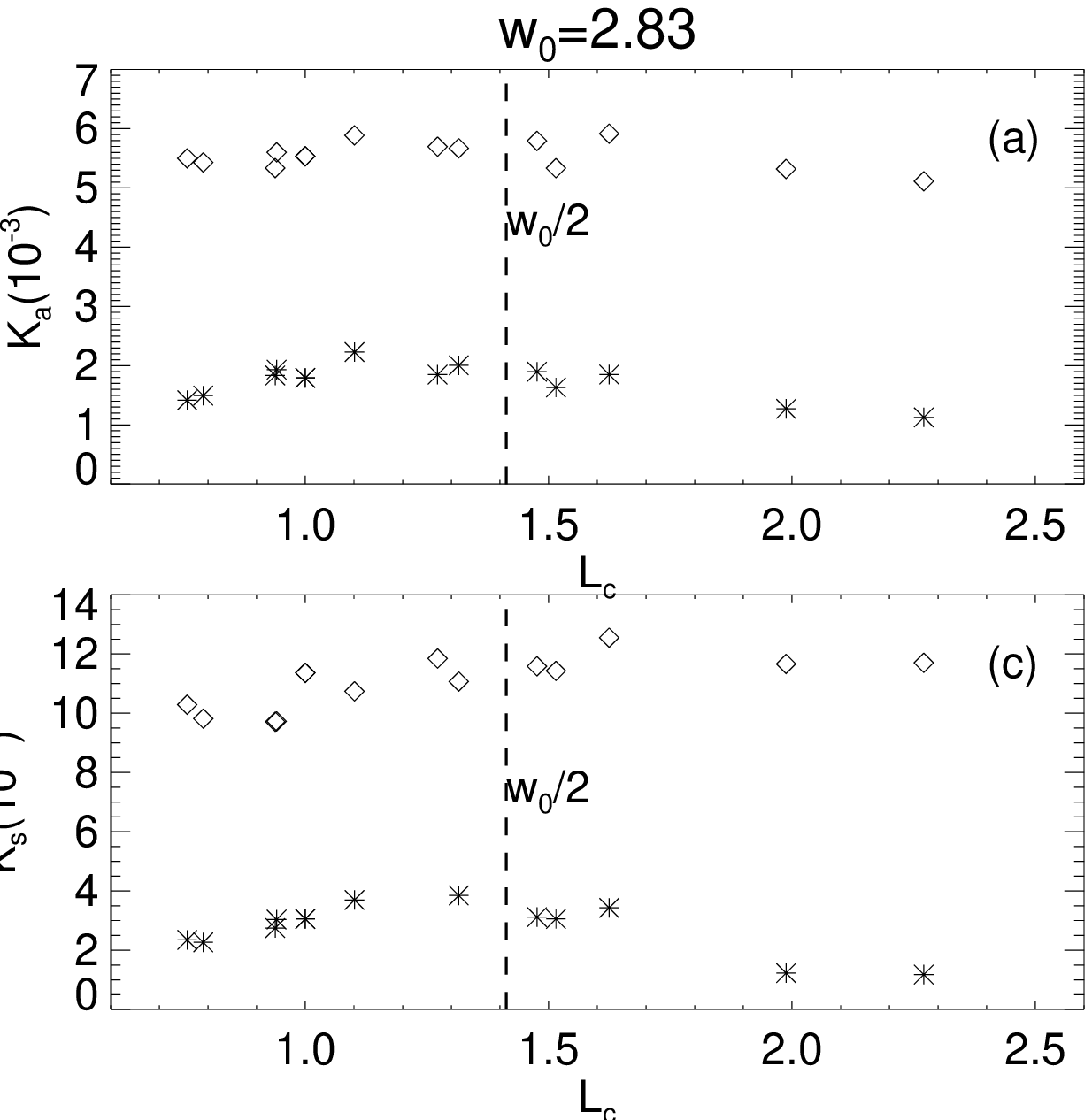}
\includegraphics[width=0.48\textwidth]{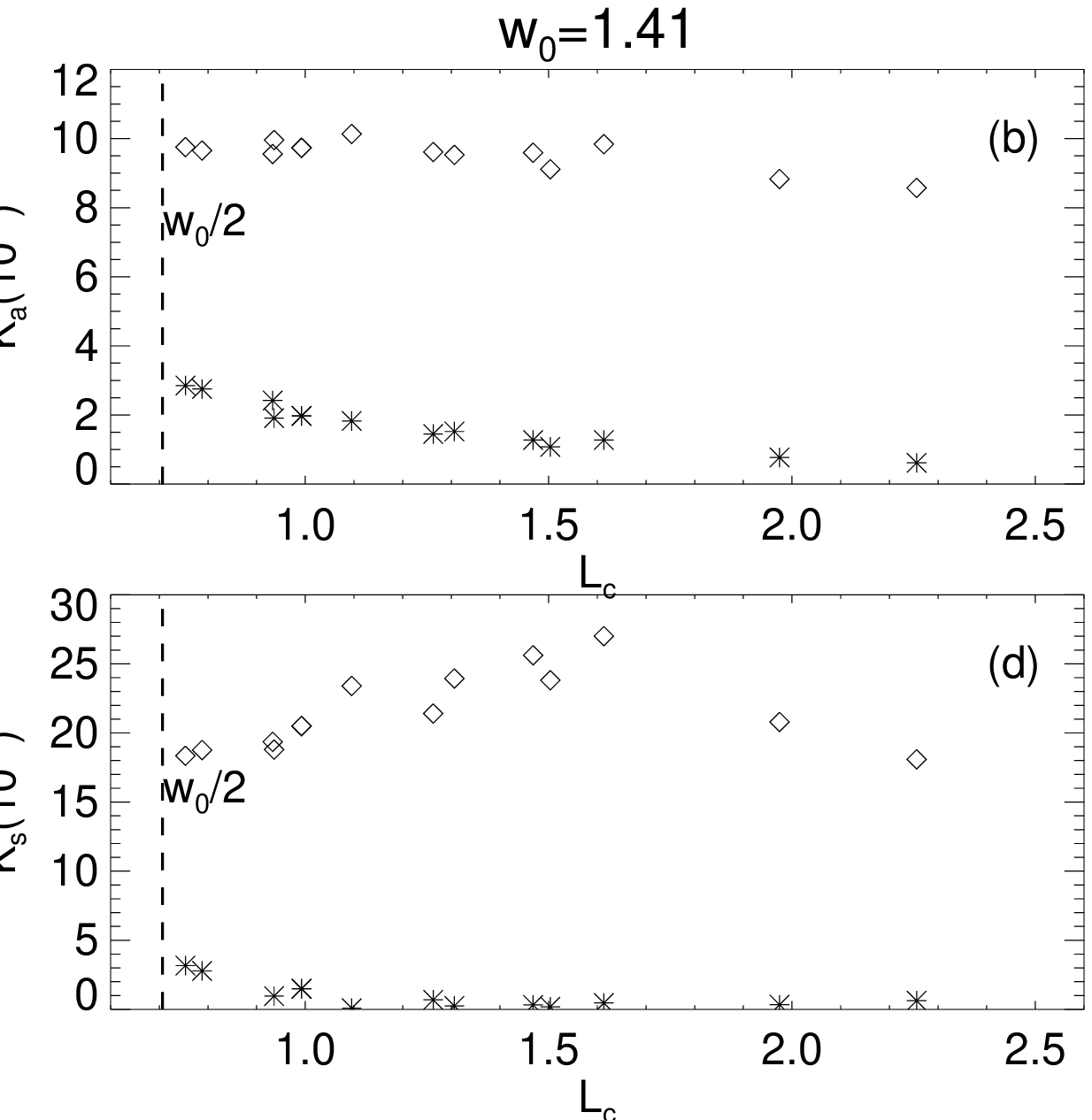}
\caption{
$K_a$ and $K_s$ as a function of the correlation length $L_c$ of the random structuring, 
    with $\delta_\rho=0.18$ for the linear (asterisks) and nonlinear (diamonds) waves. 
Two sets of pulses were launched with $w_0=2.83$ (a,c) or $w_0=1.41$ (b,d).}
\label{fig:corrlength}
\end{figure*}

Finally we examine the effect of the correlation length $L_c$ of the random structuring on the fast pulse behaviour. 
\figref{fig:corrlength}{a} and \ref{fig:corrlength}(c) show that $K_a$ and $K_s$ reach larger values for $L_c\simeq w_0/2$ in linear regime. 
This is in agreement with the results of \secref{sec:width}. 
In the case of a pulse with a sufficiently narrow initial width no clear resonance region is found (see \figref{fig:corrlength}{b} and \ref{fig:corrlength}(d)).

\section{Conclusions}
\label{sec:disc}

We performed a parametric study on fast magnetoacoustic pulses propagating in a randomly structured plasma
    perpendicular to the magnetic field. 
The fast wave is initiated with the form of a single Gaussian pulse. 
During propagation, the main pulse excites secondary propagating and standing waves, thereby losing its energy. 
Due to the random structuring, the pulse experiences attenuation and broadening. 
Thus a randomly structured plasma is effectively a dispersive medium in the context of pulse propagation. 
This effect is similar to the recently considered dispersive evolution of coronal fast wave trains \citep{yuan2013fw,pascoe2013,pascoe2014a,oliver2014,nistico2014}. 
In addition, even if the plasma is considered as being ideal, random structuring introduces effective dissipation of the wave that needs to be taken into account.

The most efficient interaction of linear low-amplitude pulses with the random structuring occurs 
    when the wavelength is about twice the correlation length of the structuring. 
Nonlinear effects cause pulse steepening which can modify this behaviour. 
Observations of this effect could be used to estimate the correlation length of the structuring in the corona. 
In general, fine structuring causes effective attenuation and dispersion for propagating pulses, and acts against nonlinear steepening.

The evolution of the amplitude and width of the fast magnetoacoustic pulse was found to be well-approximated
    by simple linear functions of the travel distance, with the exception of nonlinear large pulse amplitude
    profile which was modelled as an exponential function. 
Several factors were identified to influence the amplitude attenuation coefficient $K_a$ and the width broadening
    coefficient $K_s$. 
A higher density contrast leads to faster attenuation and broadening of wider pulses. 
For narrower pulses, splitting of the original pulse competes with dispersive broadening and so the broadening coefficient may be zero or negative.

One interesting application of our results is connected with EIT wave studies. 
In the context of EIT wave evolution, the ideal plasma of the corona could be considered as an effectively dissipative and dispersive medium. 
Many EIT wave features are reproduced by our simulations. 
The pulse profile deviates from its initial Gaussian shape \citep[see Figure~11 in][]{muhr2011}, 
    and generates an oscillatory wave, splits or develops ripple structure at later stages \citep{liu2010,shen2012}. 
The effect of nonlinear steepening is also demonstrated in our simulations \citep{veronig2010}. 
Scattering of the main pulse by non-uniformities leads to the excitation of secondary waves~\citep{li2012,shen2012,olmedo2012,shen2013,yang2013}
    and propagation dips \citep{muhr2011}. 
Our study is particularly relevant to interpreting the apparent decay of EIT waves \citep[e.g.,][]{ballai2007,long2011}, 
    which could be caused by the combination of geometrical effects (e.g., wave propagation and expansion on spherical or cylindrical surfaces), 
    non-ideal MHD effects (viscosity, resistivity, radiation and thermal conduction) and scattering and dispersive broadening
    due to the random fine structuring discussed in this study. 
The amplitude and width evolution has been measured by a number of authors, e.g., \citet{long2011,shen2012}. 
Although an apparently exponential decay in the wave amplitude was detected, 
    only a simple linear fit has been attempted \citep{long2011}. 
Our simulations suggest that an exponential fit is more appropriate in nonlinear pulses. 
The profile broadening of EIT waves could be approximated by a linear function \citep{long2011b}; 
    however, zero or negative broadening occurs in some narrow pulses of finite amplitude \citep{warmuth2004b}. 
Likewise, the analysis of nonlinear effects in EIT wave propagation \citep[e.g.][]{afanasyev2011} needs to account for
    the effective dispersion of the corona. 
Our modelling shows that a more detailed analysis of the amplitude and width evolution of EIT waves observed with
    high spatial and time resolution may allow one to fully exploit the plasma diagnostics potential.

Our model could be extended to 2D and 3D MHD models. 
Multi-dimensional MHD simulations could investigate other physics in inhomogeneous plasmas
    and will allow mode conversion \citep{sedlacek1972,ionson1978,hollweg1988,ruderman2002,goossens2002}. 
Also, line-tying boundary conditions will create fast magnetoacoustic resonators and a possibility for the excitation of standing modes
    (e.g., of the kink and sausage symmetry) in addition to the secondary propagating waves. 
Waves propagating through structures with low density contrasts as we consider here for a finely structured corona
    may demonstrate non-exponential damping behaviour \citep{pascoe2012,pascoe2013,hood2013}. 
Large local \alfven speed gradients also allow phase-mixing \citep{heyvaerts1983,similon1989,ofman2002b} which may play
    an important role in dissipating a fast pulse in a finely structured plasma. 
Other non-ideal processes may also be enhanced by the large local gradients provided by fine structuring.

\acknowledgements
This work is supported by the National Basic Research Program of China via Grant 2012CB825601 (DY and BL),
     FWO Odysseus Programme (DY), 
     the Marie Curie PIRSES-GA-2011-295272 \textit{RadioSun} project, 
     the European Research Council under the \textit{SeismoSun} Research Project No.~321141 (DY, DJP, VMN), 
     the Open Research Program KLSA201312 of Key Laboratory of Solar Activity of National Astronomical Observatories of China (DY), 
     STFC consolidated grant ST/L000733/1, 
     the BK21 plus program through the National Research Foundation funded by the Ministry of Education of Korea  (VMN), 
     the National Natural Science Foundation of China (40904047, 41174154, and 41274176), 
     and the Provincial Natural Science Foundation of Shandong via Grant JQ201212 (BL, DY).

\bibliographystyle{apj}
\bibliography{yuan}

\clearpage

\clearpage

\clearpage

\clearpage

\clearpage

\clearpage

\clearpage

\clearpage

\end{document}